\documentclass[prd,twocolumn,groupedaddress,nofootinbib,english,floatfix,superscriptaddress,preprintnumbers,notitlepage]{revtex4-1}
\usepackage{graphicx}
\usepackage{cancel}
\usepackage{amssymb}
\usepackage{textcomp}
\usepackage{amsmath}
\usepackage{mathtools}
\usepackage{bm}
\usepackage{times}
\usepackage{epsfig}
\usepackage[dvipsnames]{xcolor}
\usepackage{graphics}
\usepackage{hyperref}
\usepackage{setspace}
\usepackage{comment}
\usepackage{subcaption}
\usepackage[utf8]{inputenc}
\usepackage[english]{babel}
\usepackage{textcomp}
\usepackage{braket}
\usepackage[section]{placeins}
\usepackage{siunitx}
\usepackage{latexsym}
\usepackage{mathrsfs}

\usepackage{tikz}
\usetikzlibrary{patterns}
\usetikzlibrary{shapes.geometric}
\usetikzlibrary{decorations.pathmorphing}
\usetikzlibrary{arrows.meta}
\tikzset{snake it/.style={decorate, decoration=snake}}
\usetikzlibrary{positioning}
\usetikzlibrary{arrows,shapes,positioning}
\usetikzlibrary{decorations.markings}
\tikzstyle arrowstyle=[scale=1]
\tikzstyle directed=[postaction={decorate,decoration={markings,mark=at position .65 with {\arrow[arrowstyle]{stealth}}}}]
\tikzstyle reverse directed=[postaction={decorate,decoration={markings,mark=at position .65 with {\arrowreversed[arrowstyle]{stealth};}}}]

\tikzset{->-/.style={decoration={
  markings,
  mark=at position #1 with {\arrow{>}}},postaction={decorate}}}

\tikzset{-<-/.style={decoration={
  markings,
  mark=at position #1 with {\arrow{<}}},postaction={decorate}}}

\newcommand{\rC}{{\mathrm{C}}}
\newcommand{\rU}{{\mathrm{U}}}
\newcommand{\rI}{{\mathrm{I}}}
\newcommand{\rII}{{\mathrm{II}}}
\newcommand{\rIII}{{\mathrm{III}}}
\newcommand{\rIV}{{\mathrm{IV}}}

\newcommand{\mH}{\ensuremath{\mathcal{H}}}
\newcommand{\mHc}{\ensuremath{\mathcal{H}_c}}
\newcommand{\mCH}{\ensuremath{\mathcal{CH}}}

\newcommand{\bR}{\ensuremath{\mathbb{R}}}
\newcommand{\bS}{\ensuremath{\mathbb{S}}}
\newcommand{\ren}{\ensuremath{\mathrm{ren}}}

\newcommand{\VEV}[1]{{\langle #1 \rangle}}
\newcommand{\del}{{\partial}}
\newcommand{\td}{\ensuremath{\,\text{d}}}
\newcommand{\wlm}{\ensuremath{\omega\ell m}}

\begin{document}
\title{Long-range correlations of the stress tensor near the Cauchy horizon}

\author{Christiane Klein}
\email{christiane.klein@univ-grenoble-alpes.fr}
\affiliation{Institut f\"ur Theoretische Physik, Universit\"at Leipzig, Br\"uderstra{\ss}e 16, 04103 Leipzig, Germany}
\affiliation{Univ. Grenoble Alpes, CNRS, IF, 38000 Grenoble, France}
\affiliation{AGM, CY Cergy Paris Université, 2 av. Adolphe Chauvin 95302 Cergy-Pontoise, France}
\author{Jochen Zahn}
\email{jochen.zahn@itp.uni-leipzig.de}
\affiliation{Institut f\"ur Theoretische Physik, Universit\"at Leipzig, Br\"uderstra{\ss}e 16, 04103 Leipzig, Germany}

\begin{abstract}
We show that the stress tensor of a real scalar quantum field on Reissner-Nordstr{\"o}m-de Sitter spacetime exhibits correlations over macroscopic distances near the Cauchy horizon. These diverge as the Cauchy horizon is approached and are universal, i.e., state-independent. This signals a breakdown of the semi-classical approximation near the Cauchy horizon. We also investigate the effect of turning on a charge of the scalar field and consider the correlation of the stress tensor between the two poles of the Cauchy horizon of Kerr-de Sitter spacetime.
\end{abstract}
\maketitle

\paragraph*{Introduction}

 On Minkowski space and in the vacuum state, the correlations $\VEV{ \hat T_{\alpha \beta}(x) \hat T_{\gamma \delta}(x') }$ of stress tensor components of a massless field at space-like separated points $x$, $x'$ fall off as $d^{-8}$ with the distance $d$ (exponentially for massive fields).
 Hence, correlations of the stress tensor over macroscopic distances are negligible.
 This is one of the (not always outspoken) assumptions underlying the semi-classical Einstein equation 
 (here $\Lambda$ is the cosmological constant, $T^{\mathrm{class}}_{\mu\nu}$ the stress tensor of classical matter and $\Psi$ the state of the quantum matter)
\begin{equation}
    G_{\mu\nu} + \Lambda g_{\mu\nu}= 8\pi \left( T^{\mathrm{class}}_{\mu\nu}+\VEV{\hat T_{\mu\nu}}_\Psi \right).
\end{equation}

An example of a situation in which correlations of the stress tensor over macroscopic distances are not negligible would be a quantum superposition of two macroscopically different spatial distributions of quantum matter, also called a \emph{gravitational cat} state. Such superpositions played a crucial role in Feynman's famous Gedanken experiment \cite{ChapelHill} which was used to argue that in a consistent theory comprising gravity and quantum matter also the gravitational field must be quantized. The analysis of this and similar Gedanken experiments has been a major activity in quantum gravity research in recent years \cite{Anastopoulos:2015zta, Bose:2017nin, Marletto:2017kzi, Belenchia:2018szb, Danielson:2021egj} and the actual realization and study of gravitational cat states is a key goal of experimental studies of quantum gravity \cite{Carlesso:2019cuh, Carney:2022dku}.

In quantum field theory on curved spacetimes (QFTCS), correlations of quantum fields over macroscopic distances are crucial in inflationary cosmology \cite{Linde:1990}, where they provide the seeds of cosmological structure formation.

In the following, we show that also near the Cauchy horizon inside black holes, there are correlations of the stress tensor over macroscopic distances. These are generically of the same order as the square of the expectation value of the stress tensor, and in particular they diverge as the Cauchy horizon is approached. Moreover, this behaviour is universal, i.e., the leading divergence of the correlations is independent of the quantum state. Hence, the occurrence of gravitational cat states (defined as states with non-negligible correlations of the stress tensor over macroscopic distances) on the Cauchy horizon of a black hole is a robust prediction of QFTCS.

All stationary black hole solutions (with the exception of non-rotating, uncharged black holes) possess a Cauchy horizon in their interior. While the metric can be smoothly extended beyond it, the extension is non-unique, as is the extension of any other field subject to hyperbolic field equations. Hence, the occurrence of a Cauchy horizon signals the breakdown of predictivity. It was conjectured by Penrose \cite{Penrose:1974} that this breakdown
is not generic, i.e., under generic perturbations of the gravitational and/or matter fields, the Cauchy horizon should become singular. This \emph{strong cosmic censorship} (sCC) conjecture motivates the study of classical \cite{MR0521510, PoissonIsrael, Ori:1991zz, Mellor:1990, Mellor:1992, Brady:1998, Luk:2015qja, Hintz:2015, Dafermos:2015bzz, Dafermos:2017dbw, Cardoso:2017, Dafermos:2018tha, Dias:2018etb, Dias:2018ufh, Cardoso:2018nvb} and quantum \cite{Birrell:1978th, Hiscock:1980wr, Markovic:1994gy, Dias:2019ery, Zilberman:2019, Hollands:2019, Hollands:2020, Klein:2021, Zilberman:2021, Zilberman:2022a, Barcelo:2022gii, McMaken:2023tft} fields near a Cauchy horizon.

For our study, we mostly consider a scalar (charged or uncharged) field on Reissner-Nordstr{\"o}m-de Sitter (RNdS) spacetime, describing a static charged black hole in a spacetime with positive cosmological constant. This choice is motivated both by physical and practical considerations: From a practical point of view, RNdS has the advantage of spherical symmetry as well as having a further (cosmological) horizon at a finite radius, which simplifies the computation of the required scattering coefficients. 

But RNdS is also interesting from a conceptual point of view, as (the Christodoulou formulation \cite{Christodoulou:2008} of) sCC can be violated in that case \cite{Cardoso:2017, Dias:2018ufh, Cardoso:2018nvb}, i.e., the classical stress tensor diverges weaker than $V^{-1}$ at the Cauchy horizon. Here $V$ is a (Kruskal) coordinate with which one can smoothly extend the metric beyond the Cauchy horizon, situated at $V = 0$. In contrast, for quantum fields, one finds the leading divergence
\begin{equation}
\label{eq:T_VV_Universal}
    \VEV{ \hat T_{VV} } \sim C V^{-2}
\end{equation}
for the expectation value of the renormalized stress tensor component $\hat T_{VV}$ near the Cauchy horizon \cite{Zilberman:2019, Hollands:2019}, with a universal (state-independent) coefficient $C$ \cite{Hollands:2019, Hintz:2023pak}.

In this work, we will consider the correlations
\begin{equation}
    \Delta \hat T_{VV}(\delta \theta) := \VEV{ \hat T_{VV}(\theta) \hat T_{VV}(\theta + \delta \theta) }_{\rU} - \VEV{ \hat T_{VV} }_{\rU}^2
\end{equation}
of $\hat T_{VV}$ in the Unruh state at angular separation $\delta \theta$ near the Cauchy horizon (we used spherical symmetry to simplify the last term on the r.h.s.). We find that for the uncharged scalar field on RNdS 
\begin{equation}
\label{eq:Delta_T_VV}
    \Delta \hat T_{VV}(\delta \theta) \sim D(\delta \theta) V^{-4},
\end{equation}
with a non-negative coefficient $D(\delta \theta)$ which is i) universal (state-independent), ii) related to the coefficient $C$ of \eqref{eq:T_VV_Universal} as $\lim_{\delta \theta \to 0} D(\delta \theta) = 2 C^2$, and (for near-extremal RNdS) iii) essentially flat (independent of $\delta \theta$) except near spacetime parameters where $C$ vanishes.

This implies that the correlations of $\hat T_{VV}$ are of the same order as (the square of) its expectation value and that these strong correlations exist over macroscopic distances (the whole Cauchy horizon),
 putting the applicability of the semi-classical Einstein equation into question. Of course, also the divergence \eqref{eq:T_VV_Universal} indicates a breakdown of the semi-classical Einstein equation for $V \to 0$. However, for the present argument we do not need to consider this limit. In fact, our argument would also apply in a regime where $|\VEV{\hat{T}_{VV}}|$ is still small, but $\Delta\hat{T}_{VV}$ is of the same order of magnitude as  $\VEV{\hat{T}_{VV}}^2$ over macroscopic distances.

For the charged scalar field, one still finds \eqref{eq:Delta_T_VV}, but for increasing charge $q$ of the field $D(\delta \theta)$ is more and more localized near $\delta \theta = 0$, i.e., correlations over macroscopic distances are suppressed.
This calls into question the genericity of the result obtained for the uncharged scalar field. As realistic black holes are (essentially) uncharged but rotating, we finally consider the case of Kerr-de Sitter (KdS) spacetime. Since all fields ``couple'' to the angular momentum in a fashion quite analogous to the ``coupling'' of a charged field to the charge of the black hole, it is conceivable by analogy to the charged case that correlations over macroscopic distances are suppressed in that case. We  calculate the correlation between the north and the south pole and again find correlations which are of the same order as the (square of the) expectation value. Hence, also in this case, there are correlations over macroscopic distances near the Cauchy horizon, so that these can be considered as a generic feature of (essentially massless) quantum fields in black hole spacetimes.

To the best of our knowledge, the explicit calculation of correlations of the stress tensor (or more generally quantum fields) on black hole spacetimes has up to now been limited to  
time-like separated points \cite{Buss:2017vud}, or to toy models which neglect scattering \cite{Balbinot:2021bnp, Fontana:2023zqz, Balbinot:2023grl}. Hence, we also present the first calculation of quantum correlations at space-like separation in black hole spacetimes.

\begin{figure}
    \centering
    \includegraphics[scale=1]{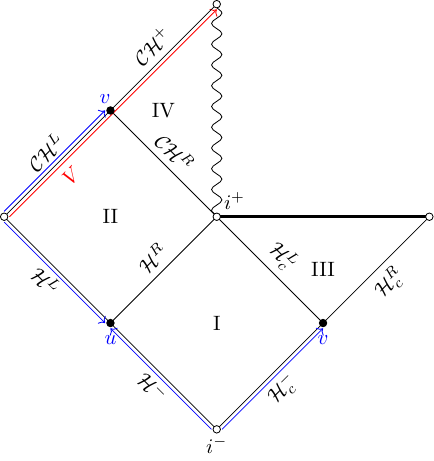}
    \caption{The Penrose diagram of RNdS. Regions $\rI$ and $\rIII$ comprise the black hole exterior. The black hole interior consists of $\rII$ and $\rIV$. The blue arrows point towards increasing $u$ and $v$, the red arrow towards increasing $V$.}
    \label{fig:PD}
\end{figure}

\paragraph*{Reissner-Nordström-de Sitter spacetime}
\label{sec:RNdS}
The RNdS spacetime is characterized by the metric
\begin{align}
\label{eq:metric}
    g&=-f(r) dt^2+f^{-1}(r)dr^2+r^2d\Omega^2\\
    f(r)&= -\frac{\Lambda}{3} r^2+1-\frac{2M}{r}+\frac{Q^2}{r^2}\, .
\end{align}
Here, $d\Omega^2$ is the area element of the unit 2-sphere, $\Lambda$ the cosmological constant, $M$ the mass, and $Q$ the charge of the black hole.
These are chosen such that $f(r)$ has three distinct positive roots $r_-<r_+<r_c$, which determine the location of the Cauchy $(\mCH)$, event $(\mH)$ and cosmological horizon $(\mHc)$. The black hole exterior consists of region $\rIII=\bR_t\times(r_c,\infty)\times\bS^2$ beyond the cosmological horizon and region $\rI=\bR_t\times(r_+,r_c)\times\bS^2$ causally connected to the black hole, as well as the horizon $\mHc^L$ connecting them. The black hole interior up to the Cauchy horizon, region $\rII=\bR_t\times(r_-,r_+)\times\bS^2$, is connected to region $\rI$ along the event horizon $\mH^R$. Region $\rII$ is connected to region $\rIV$ containing the singularity by the Cauchy horizon $\mCH^R$. A Penrose diagram of the spacetime is shown in Fig.~\ref{fig:PD}.

We introduce a double null coordinate system. 
First, we define the tortoise coordinate $r_*$ by $fdr_*=dr$. Then, one defines $u=t-r_*$ and $v=t+r_*$. These coordinates run over $\bR$ in each of the regions $\rI$, $\rII$, and $\rIII$ separately. To cover the horizons, one can introduce Kruskal-type coordinates. The null coordinate that is regular across the Cauchy horizon $\mCH^R$ is defined in region $\rII$ by
\begin{align}
    V & = -e^{-\kappa_- v}, & \kappa_i & =\frac{1}{2} \left\vert \partial_r f(r_i)\right\vert ,
    \end{align}
so $V=0$ on $\mCH^R$. These coordinates are indicated in Fig.~\ref{fig:PD}.

The other Kruskal coordinates relevant to this work are defined in region $\rI$ by $ U=-e^{-\kappa_+u}$ and $V_c=-e^{-\kappa_c v}$.

\paragraph*{Correlations of the energy flux}
\label{sec:KG field}

We will mostly consider a real, minimally coupled massive scalar field of mass $\mu^2=2\Lambda/3$, i.e., satisfying the Klein-Gordon equation
\begin{equation}
\label{eq:KGE}
    ( \nabla^\nu \nabla_\nu - \mu^2)\phi=0 .
\end{equation}
This is the same equation as for a massless, conformally coupled scalar field on RNdS (or KdS), so that we say that the field has ``conformal mass''.
For this choice the corresponding mode equation can be brought into Heun form \cite{Suzuki:1998vy} (as for massless fields of higher spin, so that $\phi$ can be seen as a proxy for the electromagnetic field), which simplifies numerical calculations.
For such a scalar field, the classical stress tensor is given by
\begin{equation}
    T_{\nu\rho} = \del_\nu \phi \del_\rho\phi - \frac{1}{2}g_{\nu\rho}\left(\del_\alpha \phi \del^\alpha\phi -\mu^2\phi^2\right) .
\end{equation}
In the quantum theory, this needs to be renormalized, since it is quadratic and local in the quantum field.
In QFTCS, this must be done in a local and covariant way \cite{Hollands:2014}, 
i.e., by Hadamard point split renormalization (up to possible finite renormalizations, which are irrelevant for the divergent behaviour near the Cauchy horizon). In this scheme, the renormalized expectation value of a Wick square with any number of derivatives is given by
\begin{align}
    &\VEV{\del^\alpha\phi(x)\del^\beta\phi(x)}^{\ren}_{\Psi} \\\nonumber
    &=\lim\limits_{x^\prime\to x}\del^\alpha_x\del^\beta_{x^\prime}\left( \VEV{\phi(x)\phi(x^\prime)}_\Psi- H(x,x^\prime)\right),
\end{align}
with $\alpha, \beta$ multi-indices.
Here, $H(x,x^\prime)$ is the Hadamard parametrix for the Klein-Gordon operator $\nabla^\nu \nabla_\nu -\mu^2$.
If the state $\Psi$ is Hadamard, i.e., with a two-point function whose singularities for $x^\prime\to x$ agree with those of $H(x,x^\prime)$, then the resulting expectation value is finite.
Thus, one may write the operator-valued distribution for the $vv$-component of the renormalized stress tensor (henceforth referred to as the \emph{energy flux}) on RNdS (somewhat formally) as
\begin{align}
\label{eq:RSET}
    \hat{T}_{vv}(x)=\lim\limits_{x^\prime\to x}\left(\del_v\phi(x)\del_v\phi(x^\prime)-\del_v\del_v^\prime H(x,x^\prime)\mathbf{1}\right).
\end{align}

We are interested in the correlations of this operator, defined by
\begin{equation}
    \Delta \hat T_{vv}(x,y)_\Psi = \VEV{\hat{T}_{vv}(x)\hat{T}_{vv}(y)}_\Psi -\VEV{\hat{T}_{vv}(x)}_\Psi\VEV{\hat{T}_{vv}(y)}_\Psi .
\end{equation}
If the state $\Psi$ is quasi-free (Gaussian), as is the case for the Unruh state considered below, then one finds, using Wick's formula,
\begin{equation}
\label{eq:Delta_T_vv}
    \Delta \hat T_{vv}(x,y)_\Psi = 2 \left(\VEV{\del_v\phi(x)\del_v\phi(y)}_\Psi \right)^2 .
\end{equation}
We would like to compute this correlation for the Unruh state on the Cauchy horizon $\mCH^R$ of RNdS.

The Unruh state for the real scalar field on RNdS is Hadamard in $\rI \cup \rII \cup \rIII$, i.e., up to the Cauchy horizon \cite{Hollands:2019}. It can be defined by expanding the quantum scalar field in terms of mode solutions to the Klein-Gordon equation \eqref{eq:KGE},
\begin{align}
    \phi(x)=\sum\limits_{\lambda,\ell, m}\int\limits_0^\infty \phi_{\wlm}^\lambda(x)a^\lambda_{\wlm}+\overline{\phi}_{\wlm}^\lambda(x)a^{\lambda\dagger}_{\wlm}\td \omega ,
\end{align}
where the $a^\lambda_{\wlm}$ and $a^{\lambda\dagger}_{\wlm}$ are creation and annihilation operators and $\phi^\lambda_{\wlm}$ form a complete set of symplectically normalized mode solutions to \eqref{eq:KGE}. The index $\lambda$ runs over two families of such modes, called ``in''- and ``up''- modes, while $\ell$ and $m$ are the usual angular quantum numbers. 

For the Unruh state, the ``up''-modes vanish near $\mHc^-\cup\mHc^R$, and are of positive frequency w.r.t.\ $U$
near $\mH^L\cup\mH^-$, while the ``in''-modes vanish near $\mH^L\cup\mH^-$ and are of positive frequency w.r.t.\ $V_c$ near $\mHc^-\cup\mHc^R$.
 
For the evaluation of the correlations at the Cauchy horizon, we will pick $x$ as any point on $\mCH^R$, and $y=x+\delta\theta$ separated from $x$ in the $\theta$-direction. Due to the spherical symmetry of the Unruh state, the correlations will only depend on $\delta\theta$.

As the Unruh state is stationary, we can also compute the correlations on $\mCH^L$, which is advantageous.
To explain this, we recall the calculation of the expectation value $\VEV{\hat T_{vv}}_{\rU}$ on $\mCH^R$ as performed in \cite{Hollands:2019}. There, a further stationary ``comparison'' state $\VEV{ \cdot }_{\rC}$ is introduced, which is defined by final data on $\mCH^L \cup \mCH^+$ (see Fig.~\ref{fig:PD}), and which is Hadamard in $\rII \cup \rIV$, i.e., across $\mCH^R$. Due to the latter property, the renormalized expectation value of $\hat T_{vv}$ in this state must vanish at $\mCH^R$, so that instead of $\VEV{\hat T_{vv}}_\rU$ one can consider $\VEV{\hat T_{vv}}_\rU - \VEV{\hat T_{vv}}_\rC$. The advantage is that the Hadamard parametrix drops out in this difference and the result can be computed (on $\mCH^L$) in terms of a mode integral as
\begin{align}
    \VEV{\hat T_{vv}}_\rU & = \sum_{\ell = 0}^\infty T^{(\ell)}_{vv}, &
\label{eq:T_ell_vv}
    T_{vv}^{(\ell)} & =\frac{2\ell+1}{16\pi^2r_-^2}\int\limits_0^\infty \td\omega\, \omega n_\ell(\omega),
\end{align}
with $n_\ell(\omega)$, explicitly given in \cite[eq. (123)]{Hollands:2019}, a function of certain scattering coefficients on RNdS.

Returning to the correlations \eqref{eq:Delta_T_vv}, we note that in principle these do not require any renormalization for spacelike separated $x$, $y$ (as is the case for $\delta \theta \neq 0$ on $\mCH^R$). However, to improve the weak convergence of the mode integral, we again use stationarity to evaluate the expression on $\mCH^L$ and subtract the correlation of $\hat T_{vv}$ in the comparison state. This does not alter the result for $\delta \theta \neq 0$, as, on $\mCH^L$, $\VEV{\del_v\phi(x)\del_v\phi(x+\delta\theta)}_\rC = 0$ for $\delta \theta \neq 0$ (a ``blind spot'' in the terminology of \cite{Levi:2016a}).
 Thus, one finds
\begin{equation}
\label{eq:dTvv}
     \Delta \hat T_{vv}(\delta\theta)_\rU = 2 \left( \sum_{\ell=0}^\infty P_\ell(\cos\delta\theta) T^{(\ell)}_{vv} \right)^2,
\end{equation}
with $P_\ell(x)$ the Legendre polynomials.
In particular, we see that the angular dependence comes from the terms with $\ell > 0$, and that $\lim_{\delta \theta \to 0} \Delta \hat T_{vv}(\delta\theta)_\rU = 2 \VEV{\hat T_{vv}}_\rU^2$.

From the result for $\Delta \hat T_{vv}(\delta\theta)_\rU$, one straightforwardly obtains (with the tensor transformation law) the divergent correlation \eqref{eq:Delta_T_VV}, with $D(\delta \theta) = \kappa_-^{-4} \Delta \hat T_{vv}(\delta \theta)_\rU$.
Furthermore, using the arguments of \cite{Hollands:2019, Hintz:2023pak}, one sees that this result is universal, i.e., the coefficient $D(\delta \theta)$ of the leading divergence is the same for all states which are Hadamard in $\rI \cup \rII \cup \rIII$.

\paragraph*{Numerical results}
\label{sec:num}
We now present numerical results for $\Delta \hat T_{vv}(\delta\theta)_\rU$. A method for the numerical computation of the integrand $n_\ell(\omega)$  with Mathematica has been developed in \cite{Hollands:2020}.
We will focus on the regime of large $Q$, since this is where sCC is violated classically. 

\begin{figure}
    \centering
    \includegraphics[scale=0.9]{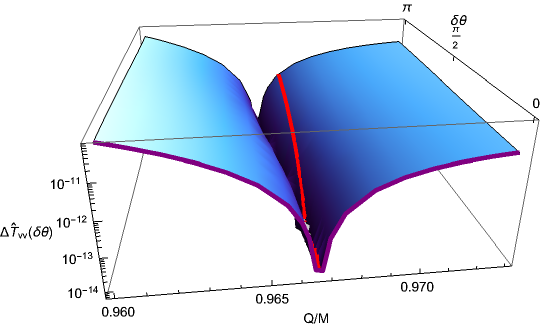}
    \caption{Correlations of the energy flux at the Cauchy horizon at angular separation $\delta\theta$ for different values of $Q/M$ and $\Lambda M^2=0.02$. The red line marks $Q_*$ at which the energy flux vanishes. The purple line represents twice the square of the flux as a function of $Q/M$. }
    \label{fig:C1}
\end{figure}

When fixing $M$ and $\Lambda$ and studying $T^{(\ell)}_{vv}$ as a function of $Q$, one finds that generically contributions with $\ell > 0$ are suppressed w.r.t.\ the $\ell = 0$ term. The only exception is a parameter region
around $Q_0$,
where $T^{(0)}_{vv}$ vanishes and changes sign.
This parameter region also contains the value $Q_*$ at which $\VEV{\hat T_{vv}}_{\rU}$ vanishes due to a cancellation of $T_{vv}^{(0)}$ and the higher $\ell$-modes, mostly $T_{vv}^{(1)}$. In Fig.~\ref{fig:C1}, we focus on a neighborhood of $Q_*$ (indicated by a red line) in parameter space. We see that away from $Q_*$ the correlations are essentially independent of $\delta\theta$, as expected due to the dominance of the $\ell = 0$ term. Furthermore, they coincide approximately with $2 \VEV{T_{vv}}_{\rU}^2$ (indicated by the purple line). 
It follows that the correlations of $\hat T_{VV}$ in the Unruh state diverge as $V^{-4}$ near the Cauchy horizon, with a coefficient of the same order as that of the leading divergence of $\VEV{\hat T_{VV}}_{\rU}^2$. We thus see strong fluctuations of the stress tensor which are correlated over macroscopic distances.

At $Q_*$, the correlations of $\hat T_{vv}$ at non-zero angular separations are non-zero and of a size similar to $\VEV{\hat T_{vv}}_{\rU}^2$ at other (nearby) values of $Q$. This means that even at $Q_*$, the typical realization of $\hat T_{vv}$ in a single measurement must be of the same order of magnitude as for other nearby values of $Q$. The positive and negative measurement results only cancel out on average. Thus, even if the leading divergence of the expectation value of the energy flux vanishes for this particular choice of parameters, one would expect that in a typical realization quantum effects will still lead to a quadratic divergence of the stress tensor and thereby restore sCC.

We note that at $Q_*$, the correlations at large angular separation are larger than at small angular separation. This is counter-intuitive, as one would usually expect the correlations to decay with the separation.

\begin{figure}
    \centering
    \includegraphics[width=0.47\textwidth]{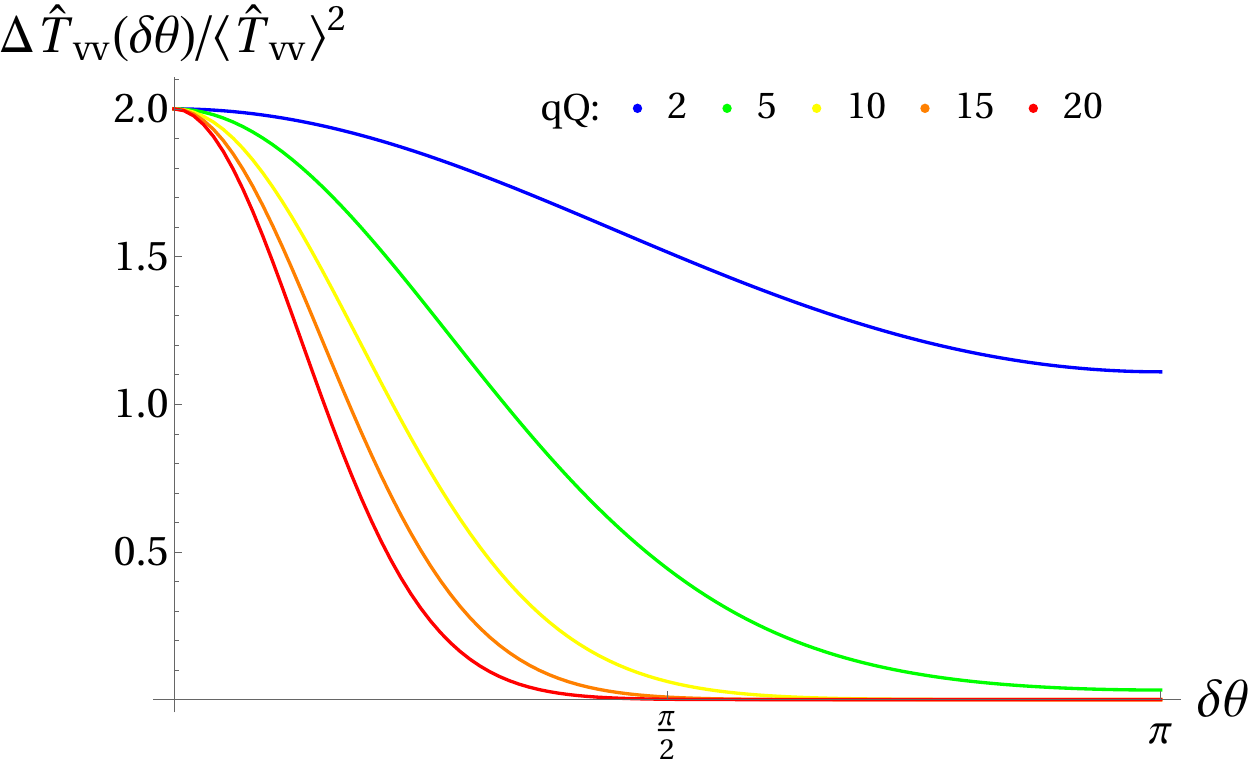}
    \caption{Correlation of the energy flux, normalized to the square of the expectation value, as a function of $\delta\theta$ for different values of the scalar field charge $q$ at $Q/M=1$ and $\Lambda M^2=0.14$. For larger $q Q$, the range of the correlations shrinks.}
    \label{fig:qdep}
\end{figure}

To see how generic our finding of strong correlations of the energy flux over macroscopic distances near the Cauchy horizon is, we study two variations of the above. The first one is to turn on a charge $q$ of the scalar field. This indeed alters the picture substantially. First, if $q$ is sufficiently large, the sign change of $T_{vv}^{(0)}$ as a function of $Q/M$ that we observed for the uncharged field is absent, and therefore also the sign change in $\VEV{\hat T_{vv}}_\rU$. Second, as $q$ is increased, the relative size of the higher-$\ell$ modes, $T_{vv}^{(\ell)}$ with $\ell\geq 1$, compared to $T_{vv}^{(0)}$, increases. This leads to a stronger dependence of the correlations on $\delta\theta$. This can be observed in Fig.~\ref{fig:qdep}, where $\Delta \hat T_{vv}(\delta \theta)_\rU$ (normalized with $\VEV{\hat T_{vv}}_\rU^2$) is shown as a function of $\delta \theta$ for different values of the field charge $q$ (for a fixed choice of spacetime parameters $\Lambda$ and $Q$). We see that as $q$ increases, the correlations start to localize stronger around $\delta\theta=0$.

\begin{figure}
    \centering
    \includegraphics[width=0.47\textwidth]{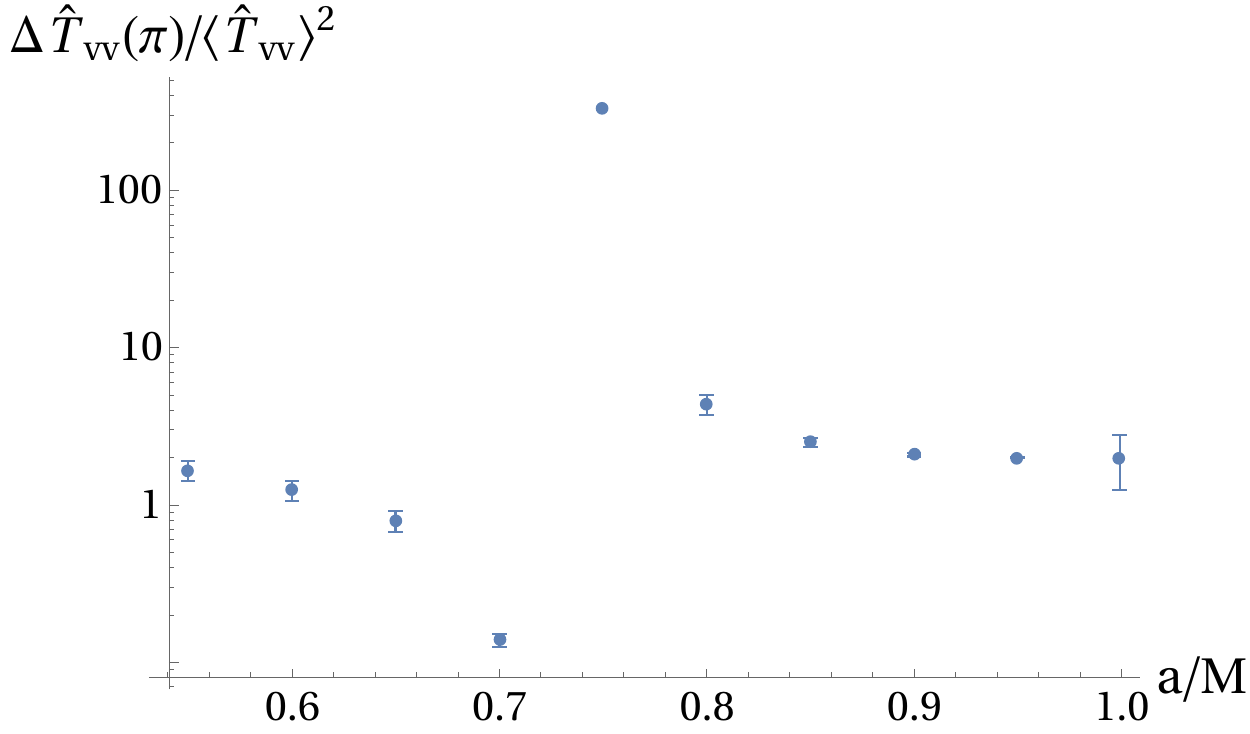}
    \caption{Correlation of the energy flux between the poles of KdS as a function of the black hole angular momentum $a/M$ for $\Lambda M^2=1/270$. The correlations are of the same order as the square of the expectation value, except near $a/M \simeq 0.75$, where $\VEV{\hat T_{vv}}(\theta = 0)$ changes its sign (for $a/M = 0.75$, the error bar is omitted, due to the small denominator).}
    \label{fig:KdS}
\end{figure}

Finally, we consider the case of a real scalar field (of conformal mass) on KdS,
which describes a rotating black hole in the presence of a positive cosmological constant. Due to the lack of spherical symmetry, this is considerably more involved than RNdS. Results for the expectation values of $\hat T_{vv}$ (and $\hat T_{v \varphi}$) on Kerr (KdS) were recently obtained in \cite{Zilberman:2022a} (\cite{Klein:2024}). For simplicity, we restrict to the correlation between the two poles.
Using the results of \cite{Kerr-dS}, one obtains the correlations shown in Fig.~\ref{fig:KdS}. 
These are of the same order as $\VEV{\hat T_{vv}}_{\rU}^2$ at the pole, except around the parameter value at which $\VEV{\hat T_{vv}}_{\rU}^2$ vanishes.
Hence, there are strong correlations over macroscopic distances, also near the Cauchy horizon of KdS. 

An intuitive understanding for this finding can easily be given: At the poles, only the $m=0$ modes are relevant, which do not couple to the angular momentum of the black hole, analogously to the real scalar field not coupling to the black hole charge; hence the similarity to the behaviour of the real scalar on RNdS. However, we also verified the existence of strong correlations (of the order of the square of the expectation value) away from the poles (for pairs of points related by reflection at the equatorial plane). In this case, the long-range correlations can be attributed to the dominance of the modes with $m=-\ell$, which are symmetric under reflection at the equatorial plane, and thus do not contribute to a suppression of the correlations for the pairs of points under consideration.

Quite generally, we expect strong correlations of the energy flux over macroscopic distances near the Cauchy horizon whenever only a small number of $\ell$- and $m$ modes contribute to the expectation value of the energy flux. This is expected to no longer hold for fields of a sizeable mass, in appropriate units: The effective potential governing the one-dimensional scattering problem for the modes of the real scalar (in the spherically symmetric case) is $V_{\mathrm{eff}} = f ( \ell (\ell+1) r^{-2} + \mu^2 + r^{-1} f')$, so that if the Compton wavelength corresponding to $\mu$ is much smaller than $r_-$, then the first ($\ell$ dependent) term is negligible w.r.t.\ the second one for a large range of $\ell$ values. Hence, in this case (which applies to astrophysical black holes and the known massive elementary particles, possibly with the exception of neutrinos), the scattering coefficients, and thus $n_\ell(\omega)$ should be essentially independent of $\ell$ for a large range of $\ell$ values, leading to a strong localization of the correlations. However, as there is at least one massless particle, the photon, the strong correlations over macroscopic distances near the Cauchy horizon persist.

\paragraph*{Conclusion}
\label{sec:sum}

We have seen that there are strong (divergent as $V^{-4}$) correlations of the component $\hat T_{VV}$ of the stress tensor near the Cauchy horizon which, in some cases, extend over macroscopic distances and are of the same order as the square 
of the expectation value. We expect this to be the generic behaviour for fields of vanishing or small mass.

If the correlations of $\hat T_{VV}$ are of the same order as the square of its expectation value over macroscopic distances, then correlations, i.e., fluctuations, can no longer be neglected, calling into question the applicability of the semi-classical Einstein equation, even if the expectation value of $\hat T_{VV}$ is still sufficiently small. We refer to \cite{Hu:2020, Pinamonti:2013} for approaches to take correlations of the stress tensor into account in the description of backreaction. Considering the results obtained in this work, an ansatz incorporating also the fluctuations of the stress tensor will be necessary to unravel the effect of quantum fields on the formation of singularities at the Cauchy horizon.

\begin{acknowledgments}
 We would like to thank M.~Soltani and M.~Casals for providing the data for the results on Kerr-de Sitter. We would also like to thank S.~Hollands and A.~Ori for stimulating discussions. This work has been funded by the Deutsche Forschungsgemeinschaft (DFG) under the Grant No. 406116891 within the Research Training Group RTG 2522/1.
\end{acknowledgments}

\bibliography{cor_bib}

\begin{thebibliography}{49}%
\makeatletter
\providecommand \@ifxundefined [1]{%
 \@ifx{#1\undefined}
}%
\providecommand \@ifnum [1]{%
 \ifnum #1\expandafter \@firstoftwo
 \else \expandafter \@secondoftwo
 \fi
}%
\providecommand \@ifx [1]{%
 \ifx #1\expandafter \@firstoftwo
 \else \expandafter \@secondoftwo
 \fi
}%
\providecommand \natexlab [1]{#1}%
\providecommand \enquote  [1]{``#1''}%
\providecommand \bibnamefont  [1]{#1}%
\providecommand \bibfnamefont [1]{#1}%
\providecommand \citenamefont [1]{#1}%
\providecommand \href@noop [0]{\@secondoftwo}%
\providecommand \href [0]{\begingroup \@sanitize@url \@href}%
\providecommand \@href[1]{\@@startlink{#1}\@@href}%
\providecommand \@@href[1]{\endgroup#1\@@endlink}%
\providecommand \@sanitize@url [0]{\catcode `\\12\catcode `\$12\catcode
  `\&12\catcode `\#12\catcode `\^12\catcode `\_12\catcode `\%12\relax}%
\providecommand \@@startlink[1]{}%
\providecommand \@@endlink[0]{}%
\providecommand \url  [0]{\begingroup\@sanitize@url \@url }%
\providecommand \@url [1]{\endgroup\@href {#1}{\urlprefix }}%
\providecommand \urlprefix  [0]{URL }%
\providecommand \Eprint [0]{\href }%
\providecommand \doibase [0]{http://dx.doi.org/}%
\providecommand \selectlanguage [0]{\@gobble}%
\providecommand \bibinfo  [0]{\@secondoftwo}%
\providecommand \bibfield  [0]{\@secondoftwo}%
\providecommand \translation [1]{[#1]}%
\providecommand \BibitemOpen [0]{}%
\providecommand \bibitemStop [0]{}%
\providecommand \bibitemNoStop [0]{.\EOS\space}%
\providecommand \EOS [0]{\spacefactor3000\relax}%
\providecommand \BibitemShut  [1]{\csname bibitem#1\endcsname}%
\let\auto@bib@innerbib\@empty
\bibitem [{\citenamefont {Feynman}(2011)}]{ChapelHill}%
  \BibitemOpen
  \bibfield  {author} {\bibinfo {author} {\bibfnamefont {R.~P.}\ \bibnamefont
  {Feynman}},\ }in\ \href@noop {} {\emph {\bibinfo {booktitle} {{The Role of
  Gravitation in Physics: Report from the 1957 Chapel Hill Conference.}}}},\
  \bibinfo {editor} {edited by\ \bibinfo {editor} {\bibfnamefont
  {D.}~\bibnamefont {Rickles}}\ and\ \bibinfo {editor} {\bibfnamefont {C.~M.}\
  \bibnamefont {DeWitt}}}\ (\bibinfo  {publisher} {{Max-Planck-Gesellschaft zur
  F\"orderung der Wissenschaften}},\ \bibinfo {address} {Berlin},\ \bibinfo
  {year} {2011})\BibitemShut {NoStop}%
\bibitem [{\citenamefont {Anastopoulos}\ and\ \citenamefont
  {Hu}(2015)}]{Anastopoulos:2015zta}%
  \BibitemOpen
  \bibfield  {author} {\bibinfo {author} {\bibfnamefont {C.}~\bibnamefont
  {Anastopoulos}}\ and\ \bibinfo {author} {\bibfnamefont {B.-L.}\ \bibnamefont
  {Hu}},\ }\href {\doibase 10.1088/0264-9381/32/16/165022} {\bibfield
  {journal} {\bibinfo  {journal} {Class. Quant. Grav.}\ }\textbf {\bibinfo
  {volume} {32}},\ \bibinfo {pages} {165022} (\bibinfo {year} {2015})},\
  \Eprint {http://arxiv.org/abs/1504.03103} {arXiv:1504.03103 [quant-ph]}
  \BibitemShut {NoStop}%
\bibitem [{\citenamefont {Bose}\ \emph {et~al.}(2017)\citenamefont {Bose},
  \citenamefont {Mazumdar}, \citenamefont {Morley}, \citenamefont {Ulbricht},
  \citenamefont {Toro\v{s}}, \citenamefont {Paternostro}, \citenamefont
  {Geraci}, \citenamefont {Barker}, \citenamefont {Kim},\ and\ \citenamefont
  {Milburn}}]{Bose:2017nin}%
  \BibitemOpen
  \bibfield  {author} {\bibinfo {author} {\bibfnamefont {S.}~\bibnamefont
  {Bose}}, \bibinfo {author} {\bibfnamefont {A.}~\bibnamefont {Mazumdar}},
  \bibinfo {author} {\bibfnamefont {G.~W.}\ \bibnamefont {Morley}}, \bibinfo
  {author} {\bibfnamefont {H.}~\bibnamefont {Ulbricht}}, \bibinfo {author}
  {\bibfnamefont {M.}~\bibnamefont {Toro\v{s}}}, \bibinfo {author}
  {\bibfnamefont {M.}~\bibnamefont {Paternostro}}, \bibinfo {author}
  {\bibfnamefont {A.}~\bibnamefont {Geraci}}, \bibinfo {author} {\bibfnamefont
  {P.}~\bibnamefont {Barker}}, \bibinfo {author} {\bibfnamefont {M.~S.}\
  \bibnamefont {Kim}}, \ and\ \bibinfo {author} {\bibfnamefont
  {G.}~\bibnamefont {Milburn}},\ }\href {\doibase
  10.1103/PhysRevLett.119.240401} {\bibfield  {journal} {\bibinfo  {journal}
  {Phys. Rev. Lett.}\ }\textbf {\bibinfo {volume} {119}},\ \bibinfo {pages}
  {240401} (\bibinfo {year} {2017})},\ \Eprint
  {http://arxiv.org/abs/1707.06050} {arXiv:1707.06050 [quant-ph]} \BibitemShut
  {NoStop}%
\bibitem [{\citenamefont {Marletto}\ and\ \citenamefont
  {Vedral}(2017)}]{Marletto:2017kzi}%
  \BibitemOpen
  \bibfield  {author} {\bibinfo {author} {\bibfnamefont {C.}~\bibnamefont
  {Marletto}}\ and\ \bibinfo {author} {\bibfnamefont {V.}~\bibnamefont
  {Vedral}},\ }\href {\doibase 10.1103/PhysRevLett.119.240402} {\bibfield
  {journal} {\bibinfo  {journal} {Phys. Rev. Lett.}\ }\textbf {\bibinfo
  {volume} {119}},\ \bibinfo {pages} {240402} (\bibinfo {year} {2017})},\
  \Eprint {http://arxiv.org/abs/1707.06036} {arXiv:1707.06036 [quant-ph]}
  \BibitemShut {NoStop}%
\bibitem [{\citenamefont {Belenchia}\ \emph {et~al.}(2018)\citenamefont
  {Belenchia}, \citenamefont {Wald}, \citenamefont {Giacomini}, \citenamefont
  {Castro-Ruiz}, \citenamefont {Brukner},\ and\ \citenamefont
  {Aspelmeyer}}]{Belenchia:2018szb}%
  \BibitemOpen
  \bibfield  {author} {\bibinfo {author} {\bibfnamefont {A.}~\bibnamefont
  {Belenchia}}, \bibinfo {author} {\bibfnamefont {R.~M.}\ \bibnamefont {Wald}},
  \bibinfo {author} {\bibfnamefont {F.}~\bibnamefont {Giacomini}}, \bibinfo
  {author} {\bibfnamefont {E.}~\bibnamefont {Castro-Ruiz}}, \bibinfo {author}
  {\bibfnamefont {{\v{C}}.}~\bibnamefont {Brukner}}, \ and\ \bibinfo {author}
  {\bibfnamefont {M.}~\bibnamefont {Aspelmeyer}},\ }\href {\doibase
  10.1103/PhysRevD.98.126009} {\bibfield  {journal} {\bibinfo  {journal} {Phys.
  Rev. D}\ }\textbf {\bibinfo {volume} {98}},\ \bibinfo {pages} {126009}
  (\bibinfo {year} {2018})},\ \Eprint {http://arxiv.org/abs/1807.07015}
  {arXiv:1807.07015 [quant-ph]} \BibitemShut {NoStop}%
\bibitem [{\citenamefont {Danielson}\ \emph {et~al.}(2022)\citenamefont
  {Danielson}, \citenamefont {Satishchandran},\ and\ \citenamefont
  {Wald}}]{Danielson:2021egj}%
  \BibitemOpen
  \bibfield  {author} {\bibinfo {author} {\bibfnamefont {D.~L.}\ \bibnamefont
  {Danielson}}, \bibinfo {author} {\bibfnamefont {G.}~\bibnamefont
  {Satishchandran}}, \ and\ \bibinfo {author} {\bibfnamefont {R.~M.}\
  \bibnamefont {Wald}},\ }\href {\doibase 10.1103/PhysRevD.105.086001}
  {\bibfield  {journal} {\bibinfo  {journal} {Phys. Rev. D}\ }\textbf {\bibinfo
  {volume} {105}},\ \bibinfo {pages} {086001} (\bibinfo {year} {2022})},\
  \Eprint {http://arxiv.org/abs/2112.10798} {arXiv:2112.10798 [quant-ph]}
  \BibitemShut {NoStop}%
\bibitem [{\citenamefont {Carlesso}\ \emph {et~al.}(2019)\citenamefont
  {Carlesso}, \citenamefont {Bassi}, \citenamefont {Paternostro},\ and\
  \citenamefont {Ulbricht}}]{Carlesso:2019cuh}%
  \BibitemOpen
  \bibfield  {author} {\bibinfo {author} {\bibfnamefont {M.}~\bibnamefont
  {Carlesso}}, \bibinfo {author} {\bibfnamefont {A.}~\bibnamefont {Bassi}},
  \bibinfo {author} {\bibfnamefont {M.}~\bibnamefont {Paternostro}}, \ and\
  \bibinfo {author} {\bibfnamefont {H.}~\bibnamefont {Ulbricht}},\ }\href
  {\doibase 10.1088/1367-2630/ab41c1} {\bibfield  {journal} {\bibinfo
  {journal} {New J. Phys.}\ }\textbf {\bibinfo {volume} {21}},\ \bibinfo
  {pages} {093052} (\bibinfo {year} {2019})},\ \Eprint
  {http://arxiv.org/abs/1906.04513} {arXiv:1906.04513 [quant-ph]} \BibitemShut
  {NoStop}%
\bibitem [{\citenamefont {Carney}\ \emph {et~al.}(2022)\citenamefont {Carney},
  \citenamefont {Chen}, \citenamefont {Geraci}, \citenamefont {M\"uller},
  \citenamefont {Panda}, \citenamefont {Stamp},\ and\ \citenamefont
  {Taylor}}]{Carney:2022dku}%
  \BibitemOpen
  \bibfield  {author} {\bibinfo {author} {\bibfnamefont {D.}~\bibnamefont
  {Carney}}, \bibinfo {author} {\bibfnamefont {Y.}~\bibnamefont {Chen}},
  \bibinfo {author} {\bibfnamefont {A.}~\bibnamefont {Geraci}}, \bibinfo
  {author} {\bibfnamefont {H.}~\bibnamefont {M\"uller}}, \bibinfo {author}
  {\bibfnamefont {C.~D.}\ \bibnamefont {Panda}}, \bibinfo {author}
  {\bibfnamefont {P.~C.~E.}\ \bibnamefont {Stamp}}, \ and\ \bibinfo {author}
  {\bibfnamefont {J.~M.}\ \bibnamefont {Taylor}},\ }in\ \href@noop {} {\emph
  {\bibinfo {booktitle} {{2022 Snowmass Summer Study}}}}\ (\bibinfo {year}
  {2022})\ \Eprint {http://arxiv.org/abs/2203.11846} {arXiv:2203.11846 [gr-qc]}
  \BibitemShut {NoStop}%
\bibitem [{\citenamefont {Linde}(1990)}]{Linde:1990}%
  \BibitemOpen
  \bibfield  {author} {\bibinfo {author} {\bibfnamefont {A.~D.}\ \bibnamefont
  {Linde}},\ }\href@noop {} {\emph {\bibinfo {title} {{Particle physics and
  inflationary cosmology}}}},\ \bibinfo {series} {Contemporary Concepts in
  Physics}, Vol.~\bibinfo {volume} {5}\ (\bibinfo  {publisher} {Harwood},\
  \bibinfo {address} {New York},\ \bibinfo {year} {1990})\ \Eprint
  {http://arxiv.org/abs/hep-th/0503203} {arXiv:hep-th/0503203} \BibitemShut
  {NoStop}%
\bibitem [{\citenamefont {Penrose}(1974)}]{Penrose:1974}%
  \BibitemOpen
  \bibfield  {author} {\bibinfo {author} {\bibfnamefont {R.}~\bibnamefont
  {Penrose}},\ }in\ \href@noop {} {\emph {\bibinfo {booktitle} {Gravitational
  Radiation and Gravitational Collapse}}},\ \bibinfo {editor} {edited by\
  \bibinfo {editor} {\bibfnamefont {C.~M.}\ \bibnamefont {DeWitt}}}\ (\bibinfo
  {publisher} {Springer},\ \bibinfo {address} {Heidelberg},\ \bibinfo {year}
  {1974})\BibitemShut {NoStop}%
\bibitem [{\citenamefont {McNamara}(1978)}]{MR0521510}%
  \BibitemOpen
  \bibfield  {author} {\bibinfo {author} {\bibfnamefont {J.~M.}\ \bibnamefont
  {McNamara}},\ }\href {\doibase 10.1098/rspa.1978.0191} {\bibfield  {journal}
  {\bibinfo  {journal} {Proc. Roy. Soc. London Ser. A}\ }\textbf {\bibinfo
  {volume} {364}},\ \bibinfo {pages} {121} (\bibinfo {year}
  {1978})}\BibitemShut {NoStop}%
\bibitem [{\citenamefont {Poisson}\ and\ \citenamefont
  {Israel}(1990)}]{PoissonIsrael}%
  \BibitemOpen
  \bibfield  {author} {\bibinfo {author} {\bibfnamefont {E.}~\bibnamefont
  {Poisson}}\ and\ \bibinfo {author} {\bibfnamefont {W.}~\bibnamefont
  {Israel}},\ }\href {\doibase 10.1103/PhysRevD.41.1796} {\bibfield  {journal}
  {\bibinfo  {journal} {Phys. Rev. D}\ }\textbf {\bibinfo {volume} {41}},\
  \bibinfo {pages} {1796} (\bibinfo {year} {1990})}\BibitemShut {NoStop}%
\bibitem [{\citenamefont {Ori}(1991)}]{Ori:1991zz}%
  \BibitemOpen
  \bibfield  {author} {\bibinfo {author} {\bibfnamefont {A.}~\bibnamefont
  {Ori}},\ }\href {\doibase 10.1103/PhysRevLett.67.789} {\bibfield  {journal}
  {\bibinfo  {journal} {Phys. Rev. Lett.}\ }\textbf {\bibinfo {volume} {67}},\
  \bibinfo {pages} {789} (\bibinfo {year} {1991})}\BibitemShut {NoStop}%
\bibitem [{\citenamefont {Mellor}\ and\ \citenamefont
  {Moss}(1990)}]{Mellor:1990}%
  \BibitemOpen
  \bibfield  {author} {\bibinfo {author} {\bibfnamefont {F.}~\bibnamefont
  {Mellor}}\ and\ \bibinfo {author} {\bibfnamefont {I.}~\bibnamefont {Moss}},\
  }\href {\doibase 10.1103/PhysRevD.41.403} {\bibfield  {journal} {\bibinfo
  {journal} {Phys. Rev. D}\ }\textbf {\bibinfo {volume} {41}},\ \bibinfo
  {pages} {403} (\bibinfo {year} {1990})}\BibitemShut {NoStop}%
\bibitem [{\citenamefont {Mellor}\ and\ \citenamefont
  {Moss}(1992)}]{Mellor:1992}%
  \BibitemOpen
  \bibfield  {author} {\bibinfo {author} {\bibfnamefont {F.}~\bibnamefont
  {Mellor}}\ and\ \bibinfo {author} {\bibfnamefont {I.}~\bibnamefont {Moss}},\
  }\href {\doibase 10.1088/0264-9381/9/4/001} {\bibfield  {journal} {\bibinfo
  {journal} {Classical and Quantum Gravity}\ }\textbf {\bibinfo {volume} {9}},\
  \bibinfo {pages} {L43} (\bibinfo {year} {1992})}\BibitemShut {NoStop}%
\bibitem [{\citenamefont {Brady}\ \emph {et~al.}(1998)\citenamefont {Brady},
  \citenamefont {Moss},\ and\ \citenamefont {Myers}}]{Brady:1998}%
  \BibitemOpen
  \bibfield  {author} {\bibinfo {author} {\bibfnamefont {P.~R.}\ \bibnamefont
  {Brady}}, \bibinfo {author} {\bibfnamefont {I.~G.}\ \bibnamefont {Moss}}, \
  and\ \bibinfo {author} {\bibfnamefont {R.~C.}\ \bibnamefont {Myers}},\ }\href
  {\doibase 10.1103/PhysRevLett.80.3432} {\bibfield  {journal} {\bibinfo
  {journal} {Phys.\ Rev.\ Lett.}\ }\textbf {\bibinfo {volume} {80}},\ \bibinfo
  {pages} {3432} (\bibinfo {year} {1998})},\ \Eprint
  {http://arxiv.org/abs/gr-qc/9801032} {arXiv:gr-qc/9801032} \BibitemShut
  {NoStop}%
\bibitem [{\citenamefont {Luk}\ and\ \citenamefont {Oh}(2017)}]{Luk:2015qja}%
  \BibitemOpen
  \bibfield  {author} {\bibinfo {author} {\bibfnamefont {J.}~\bibnamefont
  {Luk}}\ and\ \bibinfo {author} {\bibfnamefont {S.-J.}\ \bibnamefont {Oh}},\
  }\href {\doibase 10.1215/00127094-3715189} {\bibfield  {journal} {\bibinfo
  {journal} {Duke Math. J.}\ }\textbf {\bibinfo {volume} {166}},\ \bibinfo
  {pages} {437} (\bibinfo {year} {2017})},\ \Eprint
  {http://arxiv.org/abs/1501.04598} {arXiv:1501.04598 [gr-qc]} \BibitemShut
  {NoStop}%
\bibitem [{\citenamefont {Hintz}\ and\ \citenamefont
  {Vasy}(2017)}]{Hintz:2015}%
  \BibitemOpen
  \bibfield  {author} {\bibinfo {author} {\bibfnamefont {P.}~\bibnamefont
  {Hintz}}\ and\ \bibinfo {author} {\bibfnamefont {A.}~\bibnamefont {Vasy}},\
  }\href {\doibase 10.1063/1.4996575} {\bibfield  {journal} {\bibinfo
  {journal} {J.\ Math.\ Phys.}\ }\textbf {\bibinfo {volume} {58}},\ \bibinfo
  {pages} {081509} (\bibinfo {year} {2017})},\ \Eprint
  {http://arxiv.org/abs/1512.08004} {arXiv:1512.08004 [math.AP]} \BibitemShut
  {NoStop}%
\bibitem [{\citenamefont {Dafermos}\ and\ \citenamefont
  {Shlapentokh-Rothman}(2017)}]{Dafermos:2015bzz}%
  \BibitemOpen
  \bibfield  {author} {\bibinfo {author} {\bibfnamefont {M.}~\bibnamefont
  {Dafermos}}\ and\ \bibinfo {author} {\bibfnamefont {Y.}~\bibnamefont
  {Shlapentokh-Rothman}},\ }\href {\doibase 10.1007/s00220-016-2771-z}
  {\bibfield  {journal} {\bibinfo  {journal} {Commun. Math. Phys.}\ }\textbf
  {\bibinfo {volume} {350}},\ \bibinfo {pages} {985} (\bibinfo {year}
  {2017})},\ \Eprint {http://arxiv.org/abs/1512.08260} {arXiv:1512.08260
  [gr-qc]} \BibitemShut {NoStop}%
\bibitem [{\citenamefont {Dafermos}\ and\ \citenamefont
  {Luk}(2017)}]{Dafermos:2017dbw}%
  \BibitemOpen
  \bibfield  {author} {\bibinfo {author} {\bibfnamefont {M.}~\bibnamefont
  {Dafermos}}\ and\ \bibinfo {author} {\bibfnamefont {J.}~\bibnamefont {Luk}},\
  }\href@noop {} {\  (\bibinfo {year} {2017})},\ \Eprint
  {http://arxiv.org/abs/1710.01722} {arXiv:1710.01722 [gr-qc]} \BibitemShut
  {NoStop}%
\bibitem [{\citenamefont {Cardoso}\ \emph
  {et~al.}(2018{\natexlab{a}})\citenamefont {Cardoso}, \citenamefont {Costa},
  \citenamefont {Destounis}, \citenamefont {Hintz},\ and\ \citenamefont
  {Jansen}}]{Cardoso:2017}%
  \BibitemOpen
  \bibfield  {author} {\bibinfo {author} {\bibfnamefont {V.}~\bibnamefont
  {Cardoso}}, \bibinfo {author} {\bibfnamefont {J.~a.~L.}\ \bibnamefont
  {Costa}}, \bibinfo {author} {\bibfnamefont {K.}~\bibnamefont {Destounis}},
  \bibinfo {author} {\bibfnamefont {P.}~\bibnamefont {Hintz}}, \ and\ \bibinfo
  {author} {\bibfnamefont {A.}~\bibnamefont {Jansen}},\ }\href {\doibase
  10.1103/PhysRevLett.120.031103} {\bibfield  {journal} {\bibinfo  {journal}
  {Phys. Rev. Lett.}\ }\textbf {\bibinfo {volume} {120}},\ \bibinfo {pages}
  {031103} (\bibinfo {year} {2018}{\natexlab{a}})},\ \Eprint
  {http://arxiv.org/abs/1711.10502} {arXiv:1711.10502 [gr-qc]} \BibitemShut
  {NoStop}%
\bibitem [{\citenamefont {Dafermos}\ and\ \citenamefont
  {Shlapentokh-Rothman}(2018)}]{Dafermos:2018tha}%
  \BibitemOpen
  \bibfield  {author} {\bibinfo {author} {\bibfnamefont {M.}~\bibnamefont
  {Dafermos}}\ and\ \bibinfo {author} {\bibfnamefont {Y.}~\bibnamefont
  {Shlapentokh-Rothman}},\ }\href {\doibase 10.1088/1361-6382/aadbcf}
  {\bibfield  {journal} {\bibinfo  {journal} {Class. Quant. Grav.}\ }\textbf
  {\bibinfo {volume} {35}},\ \bibinfo {pages} {195010} (\bibinfo {year}
  {2018})},\ \Eprint {http://arxiv.org/abs/1805.08764} {arXiv:1805.08764
  [gr-qc]} \BibitemShut {NoStop}%
\bibitem [{\citenamefont {Dias}\ \emph {et~al.}(2018)\citenamefont {Dias},
  \citenamefont {Reall},\ and\ \citenamefont {Santos}}]{Dias:2018etb}%
  \BibitemOpen
  \bibfield  {author} {\bibinfo {author} {\bibfnamefont {O.~J.~C.}\
  \bibnamefont {Dias}}, \bibinfo {author} {\bibfnamefont {H.~S.}\ \bibnamefont
  {Reall}}, \ and\ \bibinfo {author} {\bibfnamefont {J.~E.}\ \bibnamefont
  {Santos}},\ }\href {\doibase 10.1007/JHEP10(2018)001} {\bibfield  {journal}
  {\bibinfo  {journal} {JHEP}\ }\textbf {\bibinfo {volume} {10}},\ \bibinfo
  {pages} {001} (\bibinfo {year} {2018})},\ \Eprint
  {http://arxiv.org/abs/1808.02895} {arXiv:1808.02895 [gr-qc]} \BibitemShut
  {NoStop}%
\bibitem [{\citenamefont {Dias}\ \emph
  {et~al.}(2019{\natexlab{a}})\citenamefont {Dias}, \citenamefont {Reall},\
  and\ \citenamefont {Santos}}]{Dias:2018ufh}%
  \BibitemOpen
  \bibfield  {author} {\bibinfo {author} {\bibfnamefont {O.~J.~C.}\
  \bibnamefont {Dias}}, \bibinfo {author} {\bibfnamefont {H.~S.}\ \bibnamefont
  {Reall}}, \ and\ \bibinfo {author} {\bibfnamefont {J.~E.}\ \bibnamefont
  {Santos}},\ }\href {\doibase 10.1088/1361-6382/aafcf2} {\bibfield  {journal}
  {\bibinfo  {journal} {Class. Quant. Grav.}\ }\textbf {\bibinfo {volume}
  {36}},\ \bibinfo {pages} {045005} (\bibinfo {year} {2019}{\natexlab{a}})},\
  \Eprint {http://arxiv.org/abs/1808.04832} {arXiv:1808.04832 [gr-qc]}
  \BibitemShut {NoStop}%
\bibitem [{\citenamefont {Cardoso}\ \emph
  {et~al.}(2018{\natexlab{b}})\citenamefont {Cardoso}, \citenamefont {Costa},
  \citenamefont {Destounis}, \citenamefont {Hintz},\ and\ \citenamefont
  {Jansen}}]{Cardoso:2018nvb}%
  \BibitemOpen
  \bibfield  {author} {\bibinfo {author} {\bibfnamefont {V.}~\bibnamefont
  {Cardoso}}, \bibinfo {author} {\bibfnamefont {J.~L.}\ \bibnamefont {Costa}},
  \bibinfo {author} {\bibfnamefont {K.}~\bibnamefont {Destounis}}, \bibinfo
  {author} {\bibfnamefont {P.}~\bibnamefont {Hintz}}, \ and\ \bibinfo {author}
  {\bibfnamefont {A.}~\bibnamefont {Jansen}},\ }\href {\doibase
  10.1103/PhysRevD.98.104007} {\bibfield  {journal} {\bibinfo  {journal} {Phys.
  Rev. D}\ }\textbf {\bibinfo {volume} {98}},\ \bibinfo {pages} {104007}
  (\bibinfo {year} {2018}{\natexlab{b}})},\ \Eprint
  {http://arxiv.org/abs/1808.03631} {arXiv:1808.03631 [gr-qc]} \BibitemShut
  {NoStop}%
\bibitem [{\citenamefont {Birrell}\ and\ \citenamefont
  {Davies}(1978)}]{Birrell:1978th}%
  \BibitemOpen
  \bibfield  {author} {\bibinfo {author} {\bibfnamefont {N.~D.}\ \bibnamefont
  {Birrell}}\ and\ \bibinfo {author} {\bibfnamefont {P.~C.~W.}\ \bibnamefont
  {Davies}},\ }\href {\doibase 10.1038/272035a0} {\bibfield  {journal}
  {\bibinfo  {journal} {Nature}\ }\textbf {\bibinfo {volume} {272}},\ \bibinfo
  {pages} {35} (\bibinfo {year} {1978})}\BibitemShut {NoStop}%
\bibitem [{\citenamefont {Hiscock}(1980)}]{Hiscock:1980wr}%
  \BibitemOpen
  \bibfield  {author} {\bibinfo {author} {\bibfnamefont {W.~A.}\ \bibnamefont
  {Hiscock}},\ }\href {\doibase 10.1103/PhysRevD.21.2057} {\bibfield  {journal}
  {\bibinfo  {journal} {Phys. Rev. D}\ }\textbf {\bibinfo {volume} {21}},\
  \bibinfo {pages} {2057} (\bibinfo {year} {1980})}\BibitemShut {NoStop}%
\bibitem [{\citenamefont {Markovic}\ and\ \citenamefont
  {Poisson}(1995)}]{Markovic:1994gy}%
  \BibitemOpen
  \bibfield  {author} {\bibinfo {author} {\bibfnamefont {D.}~\bibnamefont
  {Markovic}}\ and\ \bibinfo {author} {\bibfnamefont {E.}~\bibnamefont
  {Poisson}},\ }\href {\doibase 10.1103/PhysRevLett.74.1280} {\bibfield
  {journal} {\bibinfo  {journal} {Phys. Rev. Lett.}\ }\textbf {\bibinfo
  {volume} {74}},\ \bibinfo {pages} {1280} (\bibinfo {year} {1995})},\ \Eprint
  {http://arxiv.org/abs/gr-qc/9411002} {arXiv:gr-qc/9411002} \BibitemShut
  {NoStop}%
\bibitem [{\citenamefont {Dias}\ \emph
  {et~al.}(2019{\natexlab{b}})\citenamefont {Dias}, \citenamefont {Reall},\
  and\ \citenamefont {Santos}}]{Dias:2019ery}%
  \BibitemOpen
  \bibfield  {author} {\bibinfo {author} {\bibfnamefont {O.~J.~C.}\
  \bibnamefont {Dias}}, \bibinfo {author} {\bibfnamefont {H.~S.}\ \bibnamefont
  {Reall}}, \ and\ \bibinfo {author} {\bibfnamefont {J.~E.}\ \bibnamefont
  {Santos}},\ }\href {\doibase 10.1007/JHEP12(2019)097} {\bibfield  {journal}
  {\bibinfo  {journal} {JHEP}\ }\textbf {\bibinfo {volume} {12}},\ \bibinfo
  {pages} {097} (\bibinfo {year} {2019}{\natexlab{b}})},\ \Eprint
  {http://arxiv.org/abs/1906.08265} {arXiv:1906.08265 [hep-th]} \BibitemShut
  {NoStop}%
\bibitem [{\citenamefont {Zilberman}\ \emph {et~al.}(2020)\citenamefont
  {Zilberman}, \citenamefont {Levi},\ and\ \citenamefont
  {Ori}}]{Zilberman:2019}%
  \BibitemOpen
  \bibfield  {author} {\bibinfo {author} {\bibfnamefont {N.}~\bibnamefont
  {Zilberman}}, \bibinfo {author} {\bibfnamefont {A.}~\bibnamefont {Levi}}, \
  and\ \bibinfo {author} {\bibfnamefont {A.}~\bibnamefont {Ori}},\ }\href
  {\doibase 10.1103/PhysRevLett.124.171302} {\bibfield  {journal} {\bibinfo
  {journal} {Phys. Rev. Lett.}\ }\textbf {\bibinfo {volume} {124}},\ \bibinfo
  {pages} {171302} (\bibinfo {year} {2020})},\ \Eprint
  {http://arxiv.org/abs/1906.11303} {arXiv:1906.11303 [gr-qc]} \BibitemShut
  {NoStop}%
\bibitem [{\citenamefont {Hollands}\ \emph
  {et~al.}(2020{\natexlab{a}})\citenamefont {Hollands}, \citenamefont {Wald},\
  and\ \citenamefont {Zahn}}]{Hollands:2019}%
  \BibitemOpen
  \bibfield  {author} {\bibinfo {author} {\bibfnamefont {S.}~\bibnamefont
  {Hollands}}, \bibinfo {author} {\bibfnamefont {R.~M.}\ \bibnamefont {Wald}},
  \ and\ \bibinfo {author} {\bibfnamefont {J.}~\bibnamefont {Zahn}},\ }\href
  {\doibase 10.1088/1361-6382/ab8052} {\bibfield  {journal} {\bibinfo
  {journal} {Class. Quant. Grav.}\ }\textbf {\bibinfo {volume} {37}},\ \bibinfo
  {pages} {115009} (\bibinfo {year} {2020}{\natexlab{a}})},\ \Eprint
  {http://arxiv.org/abs/1912.06047} {arXiv:1912.06047 [gr-qc]} \BibitemShut
  {NoStop}%
\bibitem [{\citenamefont {Hollands}\ \emph
  {et~al.}(2020{\natexlab{b}})\citenamefont {Hollands}, \citenamefont {Klein},\
  and\ \citenamefont {Zahn}}]{Hollands:2020}%
  \BibitemOpen
  \bibfield  {author} {\bibinfo {author} {\bibfnamefont {S.}~\bibnamefont
  {Hollands}}, \bibinfo {author} {\bibfnamefont {C.}~\bibnamefont {Klein}}, \
  and\ \bibinfo {author} {\bibfnamefont {J.}~\bibnamefont {Zahn}},\ }\href
  {\doibase 10.1103/PhysRevD.102.085004} {\bibfield  {journal} {\bibinfo
  {journal} {Phys. Rev. D}\ }\textbf {\bibinfo {volume} {102}},\ \bibinfo
  {pages} {085004} (\bibinfo {year} {2020}{\natexlab{b}})},\ \Eprint
  {http://arxiv.org/abs/2006.10991} {arXiv:2006.10991 [gr-qc]} \BibitemShut
  {NoStop}%
\bibitem [{\citenamefont {Klein}\ \emph {et~al.}(2021)\citenamefont {Klein},
  \citenamefont {Zahn},\ and\ \citenamefont {Hollands}}]{Klein:2021}%
  \BibitemOpen
  \bibfield  {author} {\bibinfo {author} {\bibfnamefont {C.}~\bibnamefont
  {Klein}}, \bibinfo {author} {\bibfnamefont {J.}~\bibnamefont {Zahn}}, \ and\
  \bibinfo {author} {\bibfnamefont {S.}~\bibnamefont {Hollands}},\ }\href
  {\doibase 10.1103/PhysRevLett.127.231301} {\bibfield  {journal} {\bibinfo
  {journal} {Phys. Rev. Lett.}\ }\textbf {\bibinfo {volume} {127}},\ \bibinfo
  {pages} {231301} (\bibinfo {year} {2021})},\ \Eprint
  {http://arxiv.org/abs/2103.03714} {arXiv:2103.03714 [gr-qc]} \BibitemShut
  {NoStop}%
\bibitem [{\citenamefont {Zilberman}\ and\ \citenamefont
  {Ori}(2021)}]{Zilberman:2021}%
  \BibitemOpen
  \bibfield  {author} {\bibinfo {author} {\bibfnamefont {N.}~\bibnamefont
  {Zilberman}}\ and\ \bibinfo {author} {\bibfnamefont {A.}~\bibnamefont
  {Ori}},\ }\href {\doibase 10.1103/PhysRevD.104.024066} {\bibfield  {journal}
  {\bibinfo  {journal} {Phys. Rev. D}\ }\textbf {\bibinfo {volume} {104}},\
  \bibinfo {pages} {024066} (\bibinfo {year} {2021})},\ \Eprint
  {http://arxiv.org/abs/2105.06521} {arXiv:2105.06521 [gr-qc]} \BibitemShut
  {NoStop}%
\bibitem [{\citenamefont {Zilberman}\ \emph {et~al.}(2022)\citenamefont
  {Zilberman}, \citenamefont {Casals}, \citenamefont {Ori},\ and\ \citenamefont
  {Ottewill}}]{Zilberman:2022a}%
  \BibitemOpen
  \bibfield  {author} {\bibinfo {author} {\bibfnamefont {N.}~\bibnamefont
  {Zilberman}}, \bibinfo {author} {\bibfnamefont {M.}~\bibnamefont {Casals}},
  \bibinfo {author} {\bibfnamefont {A.}~\bibnamefont {Ori}}, \ and\ \bibinfo
  {author} {\bibfnamefont {A.~C.}\ \bibnamefont {Ottewill}},\ }\href {\doibase
  10.1103/PhysRevLett.129.261102} {\bibfield  {journal} {\bibinfo  {journal}
  {Phys. Rev. Lett.}\ }\textbf {\bibinfo {volume} {129}},\ \bibinfo {pages}
  {261102} (\bibinfo {year} {2022})},\ \Eprint
  {http://arxiv.org/abs/2203.08502} {arXiv:2203.08502 [gr-qc]} \BibitemShut
  {NoStop}%
\bibitem [{\citenamefont {Barcel\'o}\ \emph {et~al.}(2022)\citenamefont
  {Barcel\'o}, \citenamefont {Boyanov}, \citenamefont {Carballo-Rubio},\ and\
  \citenamefont {Garay}}]{Barcelo:2022gii}%
  \BibitemOpen
  \bibfield  {author} {\bibinfo {author} {\bibfnamefont {C.}~\bibnamefont
  {Barcel\'o}}, \bibinfo {author} {\bibfnamefont {V.}~\bibnamefont {Boyanov}},
  \bibinfo {author} {\bibfnamefont {R.}~\bibnamefont {Carballo-Rubio}}, \ and\
  \bibinfo {author} {\bibfnamefont {L.~J.}\ \bibnamefont {Garay}},\ }\href
  {\doibase 10.1103/PhysRevD.106.124006} {\bibfield  {journal} {\bibinfo
  {journal} {Phys. Rev. D}\ }\textbf {\bibinfo {volume} {106}},\ \bibinfo
  {pages} {124006} (\bibinfo {year} {2022})},\ \Eprint
  {http://arxiv.org/abs/2203.13539} {arXiv:2203.13539 [gr-qc]} \BibitemShut
  {NoStop}%
\bibitem [{\citenamefont {McMaken}\ and\ \citenamefont
  {Hamilton}(2023)}]{McMaken:2023tft}%
  \BibitemOpen
  \bibfield  {author} {\bibinfo {author} {\bibfnamefont {T.}~\bibnamefont
  {McMaken}}\ and\ \bibinfo {author} {\bibfnamefont {A.~J.~S.}\ \bibnamefont
  {Hamilton}},\ }\href {\doibase 10.1103/PhysRevD.107.085010} {\bibfield
  {journal} {\bibinfo  {journal} {Phys. Rev. D}\ }\textbf {\bibinfo {volume}
  {107}},\ \bibinfo {pages} {085010} (\bibinfo {year} {2023})},\ \Eprint
  {http://arxiv.org/abs/2301.12319} {arXiv:2301.12319 [gr-qc]} \BibitemShut
  {NoStop}%
\bibitem [{\citenamefont {Christodoulou}(2009)}]{Christodoulou:2008}%
  \BibitemOpen
  \bibfield  {author} {\bibinfo {author} {\bibfnamefont {D.}~\bibnamefont
  {Christodoulou}},\ }\href@noop {} {\emph {\bibinfo {title} {{The Formation of
  Black Holes in General Relativity}}}}\ (\bibinfo  {publisher} {European
  Mathematical Society Publishing House},\ \bibinfo {address} {Zürich},\
  \bibinfo {year} {2009})\ \Eprint {http://arxiv.org/abs/0805.3880}
  {arXiv:0805.3880 [gr-qc]} \BibitemShut {NoStop}%
\bibitem [{\citenamefont {Hintz}\ and\ \citenamefont
  {Klein}(2024)}]{Hintz:2023pak}%
  \BibitemOpen
  \bibfield  {author} {\bibinfo {author} {\bibfnamefont {P.}~\bibnamefont
  {Hintz}}\ and\ \bibinfo {author} {\bibfnamefont {C.~K.~M.}\ \bibnamefont
  {Klein}},\ }\href {\doibase 10.1088/1361-6382/ad2cee} {\bibfield  {journal}
  {\bibinfo  {journal} {Class. Quant. Grav.}\ }\textbf {\bibinfo {volume}
  {41}},\ \bibinfo {pages} {075006} (\bibinfo {year} {2024})},\ \Eprint
  {http://arxiv.org/abs/2310.19655} {arXiv:2310.19655 [gr-qc]} \BibitemShut
  {NoStop}%
\bibitem [{\citenamefont {Buss}\ and\ \citenamefont
  {Casals}(2018)}]{Buss:2017vud}%
  \BibitemOpen
  \bibfield  {author} {\bibinfo {author} {\bibfnamefont {C.}~\bibnamefont
  {Buss}}\ and\ \bibinfo {author} {\bibfnamefont {M.}~\bibnamefont {Casals}},\
  }\href {\doibase 10.1016/j.physletb.2017.11.048} {\bibfield  {journal}
  {\bibinfo  {journal} {Phys. Lett. B}\ }\textbf {\bibinfo {volume} {776}},\
  \bibinfo {pages} {168} (\bibinfo {year} {2018})},\ \Eprint
  {http://arxiv.org/abs/1709.05990} {arXiv:1709.05990 [hep-th]} \BibitemShut
  {NoStop}%
\bibitem [{\citenamefont {Balbinot}\ and\ \citenamefont
  {Fabbri}(2022)}]{Balbinot:2021bnp}%
  \BibitemOpen
  \bibfield  {author} {\bibinfo {author} {\bibfnamefont {R.}~\bibnamefont
  {Balbinot}}\ and\ \bibinfo {author} {\bibfnamefont {A.}~\bibnamefont
  {Fabbri}},\ }\href {\doibase 10.1103/PhysRevD.105.045010} {\bibfield
  {journal} {\bibinfo  {journal} {Phys. Rev. D}\ }\textbf {\bibinfo {volume}
  {105}},\ \bibinfo {pages} {045010} (\bibinfo {year} {2022})},\ \Eprint
  {http://arxiv.org/abs/2107.00702} {arXiv:2107.00702 [gr-qc]} \BibitemShut
  {NoStop}%
\bibitem [{\citenamefont {Fontana}\ and\ \citenamefont
  {Rinaldi}(2023)}]{Fontana:2023zqz}%
  \BibitemOpen
  \bibfield  {author} {\bibinfo {author} {\bibfnamefont {M.}~\bibnamefont
  {Fontana}}\ and\ \bibinfo {author} {\bibfnamefont {M.}~\bibnamefont
  {Rinaldi}},\ }\href@noop {} {\  (\bibinfo {year} {2023})},\ \Eprint
  {http://arxiv.org/abs/2302.08804} {arXiv:2302.08804 [gr-qc]} \BibitemShut
  {NoStop}%
\bibitem [{\citenamefont {Balbinot}\ and\ \citenamefont
  {Fabbri}(2023)}]{Balbinot:2023grl}%
  \BibitemOpen
  \bibfield  {author} {\bibinfo {author} {\bibfnamefont {R.}~\bibnamefont
  {Balbinot}}\ and\ \bibinfo {author} {\bibfnamefont {A.}~\bibnamefont
  {Fabbri}},\ }\href {\doibase 10.1103/PhysRevD.108.045004} {\bibfield
  {journal} {\bibinfo  {journal} {Phys. Rev. D}\ }\textbf {\bibinfo {volume}
  {108}},\ \bibinfo {pages} {045004} (\bibinfo {year} {2023})},\ \Eprint
  {http://arxiv.org/abs/2303.11039} {arXiv:2303.11039 [gr-qc]} \BibitemShut
  {NoStop}%
\bibitem [{\citenamefont {Suzuki}\ \emph {et~al.}(1998)\citenamefont {Suzuki},
  \citenamefont {Takasugi},\ and\ \citenamefont {Umetsu}}]{Suzuki:1998vy}%
  \BibitemOpen
  \bibfield  {author} {\bibinfo {author} {\bibfnamefont {H.}~\bibnamefont
  {Suzuki}}, \bibinfo {author} {\bibfnamefont {E.}~\bibnamefont {Takasugi}}, \
  and\ \bibinfo {author} {\bibfnamefont {H.}~\bibnamefont {Umetsu}},\ }\href
  {\doibase 10.1143/PTP.100.491} {\bibfield  {journal} {\bibinfo  {journal}
  {Prog. Theor. Phys.}\ }\textbf {\bibinfo {volume} {100}},\ \bibinfo {pages}
  {491} (\bibinfo {year} {1998})},\ \Eprint
  {http://arxiv.org/abs/gr-qc/9805064} {arXiv:gr-qc/9805064} \BibitemShut
  {NoStop}%
\bibitem [{\citenamefont {Hollands}\ and\ \citenamefont
  {Wald}(2015)}]{Hollands:2014}%
  \BibitemOpen
  \bibfield  {author} {\bibinfo {author} {\bibfnamefont {S.}~\bibnamefont
  {Hollands}}\ and\ \bibinfo {author} {\bibfnamefont {R.~M.}\ \bibnamefont
  {Wald}},\ }\href {\doibase 10.1016/j.physrep.2015.02.001} {\bibfield
  {journal} {\bibinfo  {journal} {Phys. Rept.}\ }\textbf {\bibinfo {volume}
  {574}},\ \bibinfo {pages} {1} (\bibinfo {year} {2015})},\ \Eprint
  {http://arxiv.org/abs/1401.2026} {arXiv:1401.2026 [gr-qc]} \BibitemShut
  {NoStop}%
\bibitem [{\citenamefont {Levi}\ and\ \citenamefont {Ori}(2016)}]{Levi:2016a}%
  \BibitemOpen
  \bibfield  {author} {\bibinfo {author} {\bibfnamefont {A.}~\bibnamefont
  {Levi}}\ and\ \bibinfo {author} {\bibfnamefont {A.}~\bibnamefont {Ori}},\
  }\href {\doibase 10.1103/PhysRevD.94.044054} {\bibfield  {journal} {\bibinfo
  {journal} {Phys. Rev. D}\ }\textbf {\bibinfo {volume} {94}},\ \bibinfo
  {pages} {044054} (\bibinfo {year} {2016})},\ \Eprint
  {http://arxiv.org/abs/1606.08451} {arXiv:1606.08451 [gr-qc]} \BibitemShut
  {NoStop}%
\bibitem [{\citenamefont {Klein}\ \emph {et~al.}(2024)\citenamefont {Klein},
  \citenamefont {Soltani}, \citenamefont {Casals},\ and\ \citenamefont
  {Hollands}}]{Klein:2024}%
  \BibitemOpen
  \bibfield  {author} {\bibinfo {author} {\bibfnamefont {C.}~\bibnamefont
  {Klein}}, \bibinfo {author} {\bibfnamefont {M.}~\bibnamefont {Soltani}},
  \bibinfo {author} {\bibfnamefont {M.}~\bibnamefont {Casals}}, \ and\ \bibinfo
  {author} {\bibfnamefont {S.}~\bibnamefont {Hollands}},\ }\href {\doibase
  10.1103/PhysRevLett.132.121501} {\bibfield  {journal} {\bibinfo  {journal}
  {Phys. Rev. Lett.}\ }\textbf {\bibinfo {volume} {132}},\ \bibinfo {pages}
  {121501} (\bibinfo {year} {2024})},\ \Eprint
  {http://arxiv.org/abs/2402.14171} {arXiv:2402.14171 [gr-qc]} \BibitemShut
  {NoStop}%
\bibitem [{\citenamefont {Hu}\ and\ \citenamefont {Verdaguer}(2020)}]{Hu:2020}%
  \BibitemOpen
  \bibfield  {author} {\bibinfo {author} {\bibfnamefont {B.-L.~B.}\
  \bibnamefont {Hu}}\ and\ \bibinfo {author} {\bibfnamefont {E.}~\bibnamefont
  {Verdaguer}},\ }\href {\doibase 10.1017/9780511667497} {\emph {\bibinfo
  {title} {{Semiclassical and Stochastic Gravity}: {Quantum Field Effects on
  Curved Spacetime}}}},\ Cambridge Monographs on Mathematical Physics\
  (\bibinfo  {publisher} {Cambridge University Press},\ \bibinfo {address}
  {Cambridge},\ \bibinfo {year} {2020})\BibitemShut {NoStop}%
\bibitem [{\citenamefont {Pinamonti}\ and\ \citenamefont
  {Siemssen}(2015)}]{Pinamonti:2013}%
  \BibitemOpen
  \bibfield  {author} {\bibinfo {author} {\bibfnamefont {N.}~\bibnamefont
  {Pinamonti}}\ and\ \bibinfo {author} {\bibfnamefont {D.}~\bibnamefont
  {Siemssen}},\ }\href {\doibase 10.1063/1.4908127} {\bibfield  {journal}
  {\bibinfo  {journal} {J. Math. Phys.}\ }\textbf {\bibinfo {volume} {56}},\
  \bibinfo {pages} {022303} (\bibinfo {year} {2015})},\ \Eprint
  {http://arxiv.org/abs/1303.3241} {arXiv:1303.3241 [gr-qc]} \BibitemShut
  {NoStop}%
\end{thebibliography}%

\end{document}